# A Multi-Step Process for Generating Multi-Platform User Interfaces using UIML

*Mir Farooq Ali, Manuel A. Pérez-Quiñones*
Virginia Tech, Computer Science
660 McBryde Hall
Blacksburg, VA 24061, USA
E-mail: {mfali, perez}@cs.vt.edu

*Marc Abrams*
Harmonia, Inc.
1715 Pratt Drive, Suite 3100
Blacksburg, VA 24060, USA
E-mail: marc@harmonia.com

**ABSTRACT**
There has been a widespread emergence of computing devices in the past few years that go beyond the capabilities of traditional desktop computers. These devices have varying input/output characteristics, modalities and interaction mechanisms. However, users want to use the same kinds of applications and access the same data and information on these appliances that they can access on their desktop computers. The user interfaces for these devices and platforms go beyond the traditional interaction metaphors. It is a challenge to build User Interfaces (UIs) for these devices of differing capabilities that allow the end users to perform the same kinds of tasks. The User Interface Markup Language (UIML) is an XML-based language that allows the canonical description of UIs for different platforms. We present a multi-step transformation-based framework for building Multi-Platform User Interfaces using UIML. We describe the language features of UIML that facilitate the development of multi-platform UIs, the multi-step process involved in our framework and the transformations needed to build the UIs.

**KEYWORDS:** Multi-Platform User Interfaces, Transformations, UIML, Physical Model, Logical Model.

**INTRODUCTION**
There has been a widespread proliferation of computing devices in the past few years. These computing devices have different interaction styles, input/output techniques, modalities, characteristics, and contexts of use. Dan Olsen [20], in the broader context of problems existing and emerging due to changes in the basics of interaction techniques, talks about problems in interaction emerging due to the diversity of interactive platforms. He uses the term "chaos" to describe the overall problems. He writes that computers of the future could well be wall-sized, desk-sized, palm-sized or even ear-sized. An immediate outcome of these devices of varying capabilities is the corresponding variety of interaction styles. He mentions that while it would be reasonable to expect these varying devices to have different interaction styles, it would not be acceptable to assume that they would operate independently of each other.

Mobile devices introduce an additional complexity. Peter Johnson [14] outlines four concerns regarding the HCI of mobile systems, one of which is accommodating the diversity and integration of devices, network services and applications. Stephen Brewster, *et al.* [6] talk about the problems associated with the small screen size of hand-held devices. In comparison to desktop computers, hand-held devices will always suffer from a lack of screen real estate, so new metaphors of interaction have to be devised for such devices. Some of the other problems that they mention while dealing with small devices were with navigation and presenting information. Two of the problems we encountered in our own work [3] in building UIs for different platforms were the different layout features and screen sizes associated with each platform and device.

Given this wide variety of devices, it is difficult to develop an application for multi-platform deployment without duplicating development effort. In this paper, we present a multi-step process of building multi-platform user interfaces that reduces the duplication of effort by factoring out common parts of interfaces in different levels. The different levels all use UIML as the representation language. The process presented here provides transformations that convert a UI from one level to the next.

## Terminology

Some terminology that is used in the paper is defined next. An *application* is defined to be the back end logic behind a UI that implements the interaction supported by the user interface. A *device* is a physical object with which an end-user interacts using a user interface, such as a Personal Computer (PC), a hand-held computer (like Palm), a cell-phone, an ordinary desktop telephone or a pager. A *platform* is a combination of a device or information appliance, Operating System (OS) and a toolkit. An example of a platform is a PC running Windows 2000 on which applications use the Java Swing toolkit. This definition has to be expanded in the case of HTML to include the version of HTML and the particular web browser being used. For example, Internet Explorer 5.5 running on a PC with Windows NT would be a different platform than Netscape Communicator 6.0 running on the same PC with the same OS since they implement different features of HTML differently. A related term that we will use in this paper is *family*, which indicates a group of platforms that have similar layout features.

*Rendering* is the process of converting a UIML document into a form that can be presented (e.g. through sight or sound) to an end-user, and with which the user can interact. Rendering can be accomplished in two forms: by compiling UIML into another language (e.g. WML or VoiceXML), and by interpreting UIML, meaning that a program reads UIML and makes call to an API that displays the user interface and allows interaction.

A *toolkit* is the library or markup language used by the application program to build its UI. A toolkit typically describes the widgets like menus, buttons and scrolling bars including their behavior. In the context of this paper, toolkit implies both markup languages like WML, XHTML, and VoiceXML with their sets of tags and more traditional APIs for imperative languages like Java Swing, Java AWT and Microsoft Foundation Classes for C++. An *UI element* or *widget* is the primitive building block provided by any UI toolkit for creation of UIs. A *UIML Element*, not to be confused with UI Element, is used to represent the language tags within the UIML language definition. A *Vocabulary* is the set of names, properties and associated behavior for UI elements available in any toolkit.

## RELATED WORK

The area of multi-platform UI development falls under the umbrella of what is being termed as the "variety challenges" [33]. There are new challenges for application and solution developers due to the emergence of a variety of users, a variety of devices and channels, and a variety of roles and functions. We would categorize the problem of multi-platform UI development arising due to the emergence of a variety of devices and channels. This research area is relatively new and there has not been a lot of published literature in this area. There have been some approaches towards solving this problem. Building "plastic interfaces" [7, 30] is one such method in which the UIs are designed to "withstand variations of context of use while preserving usability". This methodology uses concepts from model-based approaches that we will discuss next in building UIs.

Transcoding [4, 12, 13] is a technique used in the World Wide Web for adaptively converting web-content for the increasingly diverse kinds of devices that are being used these days to access web pages. This process also requires some kind of transformation that occurs on the HTML web page to convert it to the desired format. Transcoding assumes multiple forms. In the simplest form, semantic meaning is inferred from the structure of the web page and the page is transformed using this semantic information. A more sophisticated version of transcoding associates annotations with the structural elements of the web page and the transformation occurs based on these annotations. Another version infers semantics based on a group of web pages. Although these approaches work, they are not too extensible since it is not always possible to infer semantic information from the structural elements of web pages.

## Model-Based tools

It is useful to revisit some of the concepts behind model-based UI development tools since we feel that some of these concepts have to be utilized for generating multi-platform UIs. Model-based user interface development tools use different kinds of high-level specification of the tasks that users need to perform, data models that capture the structure and relationships of the information that applications manipulate, specifications of the presentation and dialogue, user models etc, and automatically generate some parts or the complete user interface [16]. One of the central ideas behind model-based tools is to achieve balance between the detailed control of the design of the UI and automation. Many model-based tools have been built in the late 80s and early 90s that include UIDE [25, 26], Interactive UIDE [10], HUMANOID [28], MASTERMIND [29], ITS [32], Mecano [23], and Mobi-D [24].

The central component in a model-based tool is the model that is used to represent the UI in an abstract fashion. Different types of models have been used in different systems including task models, dialogue models, user models, domain models, and application models. All these models represent the UI at a higher level of abstraction than what is possible with a more concrete representation. The UI developer built these models that were transformed either automatically or semi-automatically to generate the final UI.

Although model-based tools provide many advantages over other user interface development tools, one of the main drawbacks of these systems was that the automatically generated interfaces were not of very good quality. It was not feasible to produce good quality interfaces for even moderately complex applications from just data and task models. One more limitation of some of the earlier systems was the lack of user control over the process of UI generation. Having more user-control over the UI development process and having more usable models could rectify some of the deficiencies of the model-based approaches.

**Markup Languages and the World Wide Web**
Since the advent of the World Wide Web in the mid-90s and the emergence of eXtensible Markup Language (XML) as a standard meta-language, a number of different markup languages have emerged for creating UIs for different devices. The foremost among these are HTML [36] for desktop machines, WML [34] for small hand-held devices, and VoiceXML [31] for voice-enabled devices.

HTML can be considered a language for multi-platform development, but it only supports platforms that are in the same family i.e., primarily desktop computers. Some efforts have tried to make HTML files available in other devices (e.g. see Transcoding above) but in general HTML has remained tied to desktop computers.

The Wireless Markup Language (WML), a part of the Wireless Application Protocol (WAP), is an XML-based language designed primarily for devices with small screen-sizes and limited bandwidth, including cellular phones and pagers. WML uses the metaphor of a deck of cards to represent a UI with a user having to navigate between different cards that are grouped together like a deck. The language comprises of 35 elements, 25 of which are mandatory.

VoiceXML is a markup language for specifying interactive voice response applications. It is designed for creating audio dialogs that feature synthesized speech, digitized audio, recognition of spoken and Dual-Tone Multi Frequency key input, recording of spoken input, telephony, and mixed-initiative conversations. XML itself is a meta-language that allows the definition of other languages. There have been various other standards developed in conjunction with XML including XSLT that allows XML to be converted to other formats based on some rules.

Xforms [35] is the next generation of HTML forms, which intends to greatly enhance the capability of the current forms available in HTML. One of the goals of XForms is to provide support for hand-held, television and desktop browsers. As part of its requirements, XForms provides a separate Model, User Interface and Data Instance. This provides some level of separation between the actual data and its presentation, a desired feature of UIs that does not exist in HTML currently. XForms is designed for form-based interfaces; its model separates the data and the processing of the data from the interface but treats the interface itself as one block. This inherent assumption that the interface is form-based makes it ideal from HTML interfaces but limits the range of interfaces that a language based on XForms can describe.

There is also work being done in the areas of device-independence and mobile devices at the World Wide Web Consortium (http://www.w3c.org) to provide greater access to different devices.

While all of these markup languages have almost entirely removed the need to know toolkit, hardware, and operating system specifics for UI development, they have not made a significant contribution towards multi-platform development.

**UIML**
UIML [1, 21, 22] is a declarative XML-based language that can be used to define user interfaces. One of the original design goals of UIML is to "reduce the time to develop user interfaces for multiple device families" [2]. A related design rationale behind UIML is to "allow a family of interfaces to be created in which the common features are factored out" [1]. This indicates that the capability to create multi-platform UIs was inherent in the design of UIML itself was. However, although UIML allows a multi-platform description of UIs, there is limited commonality in the platform-specific descriptions when platform-specific vocabularies are used. This means that the UI designer would have to create separate user interfaces for each platform using its own vocabulary which is defined to be a set of user interface elements with associated properties and behavior. We now present the different language features of UIML.

One of the primary design goals of UIML is to provide a canonical format for describing interfaces that map to multiple devices. Phanouriou [21, 22] lists some of the criteria used in designing UIML to generate interfaces for multiple devices using a single canonical format:

1. UIML should be able to map the interface description to a particular device/platform

2. UIML should be able to separately describe the content, structure, behavior and style aspects of an interface

3. UIML should be able to describe the behavior in a device-independent fashion.

4. UIML should be able to give the same power to the UI implementer as the native toolkit

**Language Overview**

Since the language is XML-based, the different components of a user interface are represented through a set of tags. The language itself does not contain any platform-specific or metaphor-dependent tags. For example, there is no tag like <window> that is directly linked to the desktop metaphor of interaction. UIML uses about thirty generic tags instead. Platform-specific renderers have to be built in order to render the interface defined in UIML for that particular platform. Associated with each platform-specific renderer is a vocabulary of the language widget-set or tags that are used to define the interface in the target platform.

A skeleton UIML document is represented in Figure 1. The first line of a UIML document identifies it as a XML document. The second line gives the location of the Document Type Definition (DTD) that provides the syntactic structure that the document must conform to. Both of these are expected XML tags.

```
<?xml version="1.0" ?>
<!DOCTYPE uiml PUBLIC "-//UIT//DTD
UIML 2.0 Draft//EN" "UIML2_0f.dtd">

<uiml>
    <head>...</head>
    <interface>...</interface>
    <peers>...</peers>
    <template>...</template>
</uiml>
```

Figure 1: Skeleton of a UIML document

At the highest level, a UIML document comprises of four components: <head>, <interface>, <peers> and <template>. The <interface> is the only component that is relevant for this discussion; information on the others can be found elsewhere [21].

1. <interface>: This is the heart of the UIML document in terms of representing the actual user interface. All the UIML elements that describe the UI are present within this tag. Figure 2 illustrates the skeleton of a UIML <interface> and its contents. The four main components are:

a. <structure>: The physical organization of the interface, including the relationships between the various UI Elements within the interface, is represented using this tag. Each <structure> is comprised of different <part>s. Each part represents the actual platform-specific UI Element and is associated with a single class of UI elements. The term "class" in UIML represents a particular category of UI elements. Different parts may be nested to represent a hierarchical relationship. There might be more than one structure in a UIML document representing different organizations of the same UI.

b. <style>: The style contains a list of properties and values used to render the interface. The properties are usually associated with individual parts within the UIML document through the part-names. Properties can also be associated with particular classes of parts. Typical properties associated with parts for Graphical User Interfaces (GUIs) could be the background color, foreground color, font, etc. It is also possible to have multiple styles within a single UIML document to be possibly associated with multiple structures or even the same structure. This facilitates the use of different styles for different contexts

c. <content>: This represents the actual content associated with the various parts of the interface. A clean separation of the content from the structure is useful when different content is needed under different contexts. This feature of UIML is very helpful when creating interfaces that might be used in multiple languages. An example of this is a UI in French and English, for which separate content is needed.

d. <behavior>: The behavior of an interface is specified by enumerating a set of conditions and associated actions within rules. UIML permits two types of conditions. The first condition is when an event occurs, while the second is true when an event occurs and the value of some data associated with the event is equal to a certain value. There are four kinds of actions that occur. The first action is to assign a value to a part's property. The second action is to call an external function or method. The third is to fire an event and the fourth action is to restructure the interface.

```
<interface>
    <structure>...</structure>
    <content>...</content>
    <behavior>...</behavior>
    <style>...</style>
</interface>
```

Figure 2: Skeleton of a UIML interface

An extremely detailed discussion about the language features can be found in the UIML language specification [21], while a discussion of the language design issues can be found in Phanouriou's dissertation [22].

Currently, there are platform-specific renderers available for UIML for a number of different platforms. These include Java, HTML, WML, VoiceXML and Palm OS. Each of these renderers has a platform-specific vocabulary associated with it to describe the UI elements, their behavior and layout. The mechanism that is currently employed for creating UIs with UIML is that the UI developer uses the platform-specific vocabulary to create a UIML document that is rendered for the target platform. These renderers are available from http://www.harmonia.com. Work started on UIML and these renderers at the Center for Human-Computer Interaction, Virginia Tech in 1999. There has been a conference devoted to UIML in March 2001 in France.

There is a great deal of differences in the vocabularies associated with each language. Consequently, the UI developer would have to learn each different vocabulary in order to build UIs that would work across multiple platforms. Using UIML as the underlying language for all these multiple platforms reduces the effort needed in comparison to the effort needed if the UIs had to be built using the native language and toolkit for each platform.

However, UIML alone cannot solve the problem. The differences between platforms, layouts, were found to be too significant to simply create one UIML file for one particular platform and expect it to be rendered on a different platform with a simple change in the vocabulary. A more abstract layer was found necessary based on our experience in creating a variety of UIs for different platforms. We had to basically redesign the complete UI due to the differences in the platform vocabularies and layouts.

**UIML FRAMEWORK FOR MULTI-PLATFORM USER INTERFACES**

The concept of building multi-platform UIs is relatively new. To envision the development process, we followed an existing approach from the usability engineering literature. One such approach [11] identifies three different phases in the UI development process: *interaction design, interaction software design* and *interaction software implementation.* Interaction design is the phase of the usability engineering cycle in which the "look and feel" and behavior of an UI is designed in response to what a user hears, sees or does. In current Usability Engineering practices, this phase is highly platform-specific.

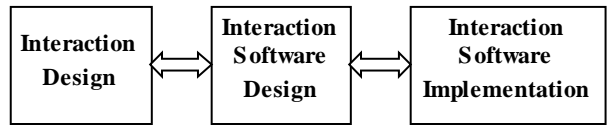

Figure 3: Traditional Usability Engineering Process for one platform

Once the interaction design is done, the interaction software is designed, which involves making decisions about the UI toolkit(s), widgets, positioning of widgets, colors, etc. Once this design is done, the software is implemented. This process of UI development for a single platform is illustrated in Figure 3.

This is the traditional view of interaction design that is highly platform-specific and works well when designing for a single platform. However, this stage of interaction design has to be further split into two distinct phases for multiple platforms: a platform-independent interaction design and platform-dependent interaction design. This phase will lead to different platform-specific interaction software designs, which in turn lead to platform-specific UIs. This process is illustrated in Figure 4.

**UIML Process**

We have developed our framework that is very closely related to the UE process discussed in Figure 4. This is illustrated in Figure 5. The main building blocks within this framework are the *Logical Model, Physical Model* and the *Platform-Specific UI*. There is also a process of transformations needed between the various models. More specifically, the logical model has to be transformed to the physical model and the physical model has to be transformed to the Platform-specific UI that is represented in UIML. Each of these building blocks is described next and the transformation process is explained. Each of the building blocks in our framework has a link to the usability engineering process that we described earlier.

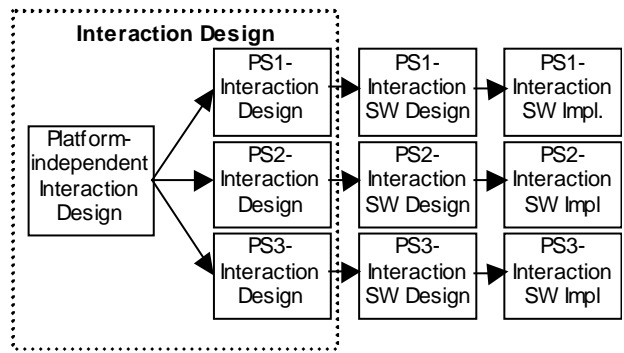

Figure 4: Usability Engineering process for multiple platforms

**Logical Model**

The main objective of the logical model is to capture the UI description at a higher level of abstraction than is possible by any physical model. The purpose of the logical model is to capture the essence of the UI so that it can be mapped to different platforms. The logical model is a collection of logical constructs that represent the UI in some abstract fashion. We are considering using a hybrid task/domain model that can be transformed to the physical model.

**Physical Model and Generic Vocabulary**

The physical model within the framework is a generic description of an UI in UIML that would work on multiple platforms. In Figure 5, we can observe that there might be more than one physical model. Each physical model is supposed to represent a group of platforms that have similar characteristics. The defining characteristic that we use in identifying different physical models is the physical layout of the UI elements on that set of platforms. For example, different HTML browsers and the Java Swing platform can all be part of one physical model based on their layout facilities. Some platforms might require a physical model of their own. The VoiceXML platform is one such example, since it is used for voice-based UIs and there is no other analogous platform for either auditory or graphical UIs.

The physical models that can be built currently are for the desktop platform (Java Swing and HTML) and phone (WML). These physical models are based on the available renderers. The specification for the physical model is already built.

Building a physical model requires building a *generic vocabulary* of UI elements, used in conjunction with UIML that can describe any UI for any platform. The advantage of UIML is apparent here since it is allows any vocabulary to be attached with it. In this case, we use a generic vocabulary that can be used in the physical model.

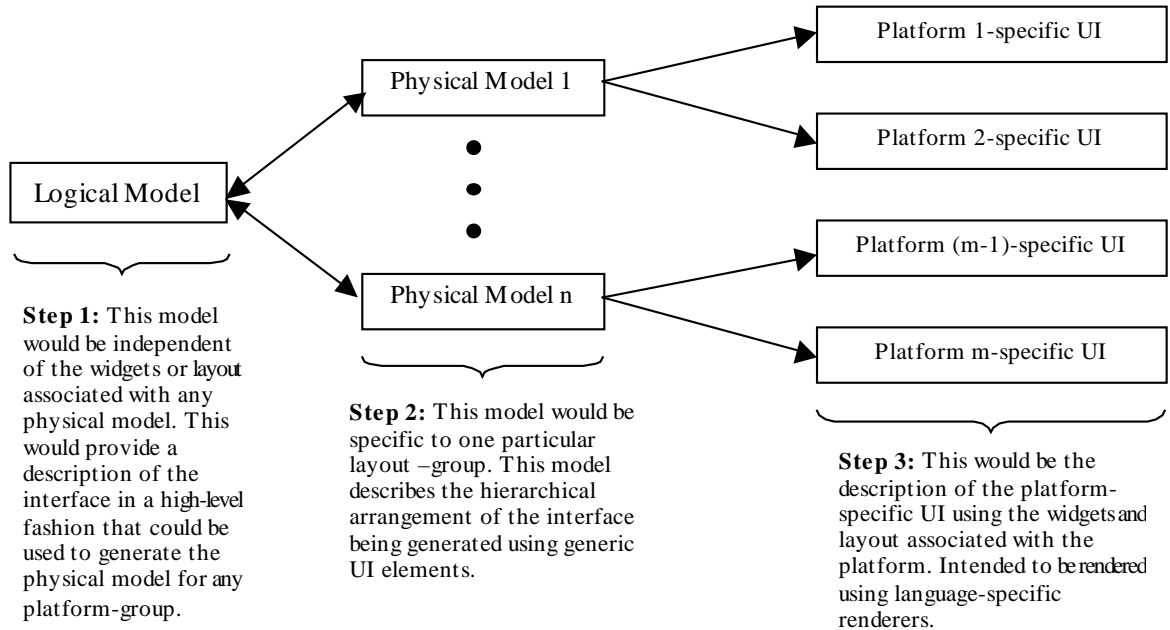

Figure 5: The overall framework for building multi-platform UIs using UIML.

A generic vocabulary can be defined to be one vocabulary for all platforms. Creating a generic vocabulary can solve some of the problems outlined above.

The vocabulary has two objectives: first, to be powerful enough to accommodate a family of devices, and second, to be generic enough to use without having expertise in all the various platforms. As a first step, a set of elements had to be selected from the platform-specific element sets. After this several generic names had to be selected that represent various user interface elements in different platforms. Next, properties and events had to be assigned for the generic elements. We have identified and defined a set of generic UI elements including their properties and events. Ali, *et al.* [3] provides a more detailed description of the generic vocabulary.

**Transformations**

There are two different types of transformations that are needed here. The first type of transformation is the mapping from the logical model to the physical model. This type of transformation has to be developer-guided and cannot be fully automated. By allowing the UI developer to intervene in the transformation and mapping process, it is possible to ensure usability. One of the main problems of some of the earlier model-based systems was that a large part of the UI generation process from the abstract models was fully automated, removing user-control of the process. Another way of describing this problem is the "mapping problem" as described by Puerta [24]. We want to eliminate this problem by having the user guide the mapping process. Once the user had identified the mappings, the system will generate the physical models based on the target platforms and the user mappings.

The second type of transformation occurs between the physical model and the platform-specific UIML. This is basically a conversion from generic UIML to platform-specific UIML, both of which could be represented as trees since they are XML-based. This process can be largely automated. However, there are certain aspects to the transformation that need to be guided by the user. For example, we have certain UI elements in our generic vocabulary that could be mapped to one or more out of a set of elements in the target platform. The developer has to select what the mapping would be for the particular target platform. The way this is currently implemented, the developer does this as a special property of the UI element.

In the next section, we present a sample application and the associated generic UIML for it.

**SAMPLE APPLICATION**

Figure 6 and Figure 7 show a sample form on the Java Swing and HTML platforms. The generic UIML that generated the two UIs is given below. For economy of space, we have not included the <style> and <behavior> sections of the UIML document. See Ali [3] for other UIML examples.

```
<?xml version="1.0"?>
<!DOCTYPE uiml PUBLIC "-//UIT//DTD UIML 2.0
  Draft//EN" "UIML2_0d.dtd">
<uiml>
 <head>
  <meta name="Purpose"
        content="Data Collection Form"/>
  <meta name="Author"
        content="Farooq Ali"/>
 </head>
<interface name="DataCollectionForm">
 <structure>
  <part name="RequestWindow"
        class="GTopContainer">
   <part name="EBlock1" class="GArea">
    <part name="TitleLabel" class="GLabel"/>
    <part name="FirstName" class="GLabel"/>
    <part name="FirstNameField" class="GText"/>
    <part name="LastName" class="GLabel"/>
    <part name="LastNameField" class="GText"/>
    <part name="StreetAddress" class="GLabel"/>
    <part name="StreetAddressField"
          class="GText"/>
    <part name="City" class="GLabel"/>
    <part name="CityField" class="GText"/>
    <part name="State" class="GLabel"/>
    <part name="StateChoice" class="GList"/>
    <part name="Zip" class="GLabel"/>
    <part name="ZipField" class="GText"/>

    <part name="OKBtn" class="GButton"/>
    <part name="CancelBtn" class="GButton"/>
    <part name="ResetBtn" class="GButton"/>
   </part>
  </part>
 </structure>
```

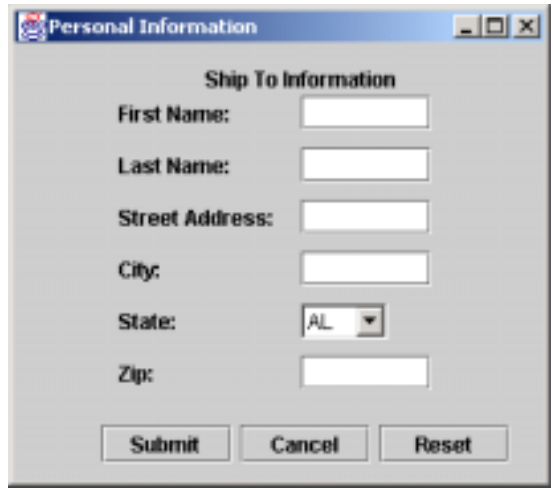

Figure 6: Screen-shot of a sample form in Java Swing.

We can see from the snippet of code above that the generic vocabulary comprises of UI elements with names that are applicable in multiple platforms. As we mentioned earlier, Java Swing and HTML comprise one physical model. So the code shown above represents just one physical model.

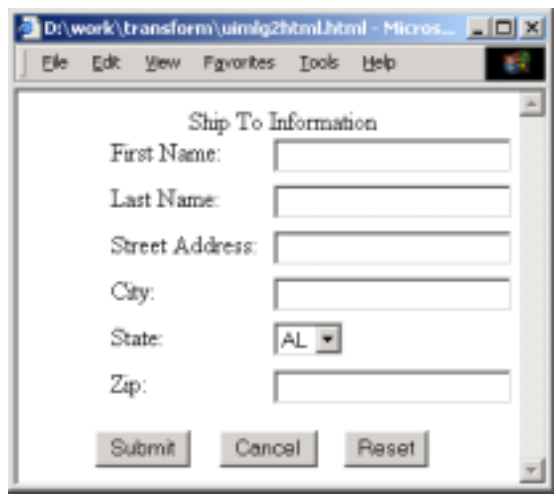

Figure 7: Screen-shot of a sample form in HTML.

It is apparent from the screen-shots that the two UIs look quite similar. In building the UI, the developer had to specify what the mappings would be for the generic UI elements like GArea and GButton for the specific platforms, since these elements could have multiple target platforms.

**CONCLUSIONS**

We have currently build the specifications for the physical model including the generic vocabulary and the transformation algorithms needed to convert a generic UIML document into a platform-specific UIML document. We are in the process of building the specification for the logical model and the transformation algorithm to convert the logical model to the physical model. Once the logical model is built, we intend to extend it in helping build accessible interfaces that could be treated as a separate platform.

We have developed a multi-step transformation based framework using the UIML language that can be used to generate multi-platform UIs. The current framework utilizes concepts from the model-based UI development literature and Usability engineering realm and applies them to this new area of multi-platform UI development. This framework tries to eliminate some of the pitfalls of the model-based approaches by having multiple steps and allowing for developer intervention throughout the UI generation process.

The physical model and the transformation algorithm that converts the physical model to the platform-specific UIML has already been developed at Harmonia, Inc, and will be incorporated within a multi-platform Authoring Tool for UIML. The logical model will be incorporated within this tool in the future.